\documentclass[12pt]{article}
\usepackage{latexsym}
\usepackage{graphicx}
\thicklines                         % thick straight lines for pictures (LaTeX)
\long \def \blockcomment #1\endcomment{}
\begin{document}           % End of preamble and beginning of text.

\baselineskip=0.33333in
%\begin{quote} \raggedleft TAUP 2846-2007
%\end{quote}
%\title{A Sample Document}  % Declares the document's title.
%\author{Leslie Lamport}    % Declares the author's name.
%\date{December 12, 1984}   % Deleting this command produces today's date.
\vglue 0.5in
%\maketitle                 % Produces the title.

\begin{center}{\bf On the Crucial Significance of the Multi-Configuration \\
Structure of a Bound State of Several Dirac Particles}
\end{center}
\begin{center}E. Comay$^*$
\end{center}

\begin{center}
Charactell Ltd. \\
P.O. Box 39019, Tel Aviv 61390 \\
Israel
\end{center}
\vglue 0.5in
\vglue 0.5in
\noindent
%\vglue 0.2in
\vglue 0.5in
\noindent
Abstract:

The structure of a bound state of several Dirac particles is discussed.
Relying on solid mathematical arguments of the Wigner-Racah algebra,
it is proved the a non-negligible number of configurations is
required for a description of this kind of systems. At present, the
main results are not widely known and this is the underlying reason for the
phenomenon called the proton spin crisis.

%\vglue 0.5in
%\noindent

\newpage
\noindent
{\bf 1. Introduction}
\vglue 0.33333in

\begin{center}
 {\em
Once upon a midnight dreary, while I pondered weak and weary,\\
Over many a quaint and curious volume of forgotten lore...}[1].
\end{center}

The objective if this work is to prove that the
multi-configuration structure of a bound state of several Dirac
particles plays an extremely important role.
The existence of such a multi-configuration structure
was already known many decades ago [2,3] and early electronic
computers were used for providing a numerical proof of this issue [4].
(Note that the first edition of [2] was published in 1935.)
Unfortunately, this scientific evidence has not found
its way to contemporary
textbooks of physics [5] and has become a kind of a forgotten lore. The
paper proves the main points of this issue and shows its far reaching
meaning and its relevance to physical problems that are still unsettled.
In doing so the paper aims to make a
contribution to the correction of this situation.

It is well known that quantum mechanics explains the Mendeleev periodic table
of chemical elements. The shell structure of electrons provides an
easy interpretation
of chemical properties of noble gases (a full shell), halogens
(a full shell minus 1), alkali metals (a full shell + 1) etc. The standard
explanation of the Mendeleev periodic table uses a single configuration for
a description of the electronic states of each chemical element. Thus,
for example, the helium and the lithium atoms are
described by the $1s^2$ and $1s^22s$ configurations, respectively.
At this point the following problem arises: Does the unique configuration
structure of an atomic ground state make an acceptable
description of its quantum mechanical system or is it just a useful
pedagogical explanation of the Mendeleev periodic table? The answer to
this problem certainly must be obtained from a mathematical analysis
of the quantum mechanical state of systems that contain
more than one electron. By describing an outline of
this task, the present work proves
beyond any doubt that an atomic state of more than one electron has
a multi-configuration structure and that no single configuration
dominates the system.

The conclusion stated above has two important aspects. First,
It is clear that a
correct understanding of the structure of any fundamental
physical system is a vital theoretical asset for
every physicist. Next,
it turns out that the lack of an adequate
awareness of this physical evidence has already
caused the phenomenon called the ``proton spin crisis" [6] which haunts
the community for decades. The measurements published
in [6] show that quarks carry a very small portion of the proton's spin
and this evidence has been regarded as a surprise.
Now, it is shown in this work that the multiconfiguration structure
found in
atomic states is not a specific property of the Coulomb interaction.
Thus, it is expected to be also found in any bound state of three
spin 1/2 quarks, like it is found in bound states of several
spin 1/2 electrons.
For this reason, one can state that if the experiment described in [6]
would have shown that {\em quarks carry the entire proton's spin then
this result should have been regarded as a real crisis of fundamental
quantum mechanical principles.}

In this work, units where $\hbar = c = 1$ are used.
The second section contains a brief description of
the main properties of a bound state
of several Dirac particles that are required for the discussion.
The underlying mathematical reasons for the multiconfiguration
structure of states are discussed in the third section. Some
aspects of the results are pointed out in the last section.

\vglue 0.66666in
\noindent
{\bf 2. General Arguments}
\vglue 0.33333in

The main objective of this work is to find a reliable mathematical
method for describing the ground state of a
bound system of spin 1/2 particles. Applying Wigner's analysis
of the Poincare group [7,8], one concludes that the total mass
(namely, energy) and the total spin are
good quantum numbers. Thus, one assumes that an energy operator
(namely, a Hamiltonian) exists. For this reason, one can construct
a Hilbert space of functions that can be used for describing the
given system as an eigenfunction of the Hamiltonian. Evidently, in
the system's rest frame, an energy eigenfunction has the time
dependent factor $\exp (-iEt)$. This factor can be removed and
the basis of the Hilbert space contains time independent functions.

The fact that every relatively stable state
has a well defined total spin $J$ can be used
for making a considerable simplification of the problem. Thus, one
uses a basis for the Hilbert space that is made of functions that
have the required spin $J$ and ignores all functions that do not satisfy
this condition. Evidently, a smaller Hilbert space
reduces the amount of
technical work needed for finding the Hamiltonian's eigenfunctions.
An additional argument holds for systems whose state is determined
by a parity conserving interaction, like the strong and the
electromagnetic interactions. Thus, one can use functions that have
a well defined parity and build the Hilbert space only from functions that
have the required parity. This procedure makes
a further simplification of the problem.

The notion of a configuration of a system of several Dirac particles is a
useful mathematical tool that satisfies the two requirements stated
above (see [2], p. 113 and [9], p. 245).
A configuration is written in the form of a product of single particle
wave functions describing the corresponding radial and orbital state
of each particle belonging to the system (the $m$ quantum number is
ignored). For atomic systems a non-relativistic notation is
commonly used and
the values of the $nl$ quantum numbers denote a configuration, like
$1s^22s^1$. In relativistic cases the variables $nlj$ (see [9], p. 245)
are used. In the latter case, the variables $nj^\pi$ (here $\pi $
denotes parity and it takes the values $\pm 1$) is an equivalent
notation for a relativistic configuration because $l=j\pm 1/2$ and
the numerical parity of the $l$-value of a Dirac spinor
upper part defines the single particle's parity.
(This work uses the $nj^\pi$ notation.)
Evidently, any acceptable configuration must be
consistent with the Pauli exclusion principle.

For any given state where the total spin $J$ and parity are given, one
can use configurations that are consistent with $J$ and the product of
the single-particle parity equals the parity of the system. The total
angular momentum $J$ is obtained from an application of the law of
vector addition of angular momentum (see [2], p. 56 and [9], p. 95).
Here the triangular condition holds (see [9], p. 98). Thus, for
example, an acceptable configuration for the two-electron
$0^+$ ground state of the helium atom must take the form
$n_1 j_1^{\pi_1} \,n_2 j_2^{\pi _2}$, where $ j_1 =j_2$ and $\pi_1 = \pi _2$.
Similarly, a description of a 2-electron state where $J^\pi = 3^+$
{\em cannot} contain a configuration of the form
$n_1{\textstyle \frac {1}{2}}^+ \,n_2{\textstyle \frac {3}{2}}^+$,
because the two $J$ values 1/2 and 3/2 can only yield a total $J=1$ or $J=2$.

At this point the structure of the relevant Hilbert space is known. It
is made of configurations that satisfy certain requirements. This
is one of the useful properties of using configurations - the relevant
Hilbert space is smaller because many configurations can be ignored
due to the total spin and parity requirements. Obviously, a smaller Hilbert
space means shorter computational efforts. Thus,
the framework needed for the analysis is established. The problem
of finding how many configurations are required for an acceptable
description of an atomic state is discussed in the following section.

\vglue 0.66666in
\noindent
{\bf 3. The Multi-Configuration Structure of Atomic States}
\vglue 0.33333in

The purpose of this section is to outline a proof that shows why
a bound state of several electrons takes the form of a linear combination of
configurations. For this purpose, the Hamiltonian matrix is constructed
for a Hilbert space whose basis is made of functions that
take a configuration form. Evidently, non-vanishing
off-diagonal matrix elements prove that the required state is a
linear combination of configurations. It is shown that this property
holds even for
the simplest atomic state of more than one electron, namely
the $J^\pi = 0^+$ ground state of the 2-electron Helium atom.

As stated in the previous section, the required Hilbert space contains
functions that have the given total spin and parity. The form
of a two electron function is written as follows
\begin{equation}
\chi({\bf r}_1,{\bf r}_2) =
F_i(r_1)F_k(r_2)(j^{\pi_1}_1j^{\pi_2}_2 JM).
\label{eq:J1J2JM}
\end{equation}
Here, $F_i(r_1),\,F_k(r_2)$ denote radial functions of the
corresponding electron,
$j_1,\,j_2,\pi _1,\pi _2$ denote the single particle spin
and parity of the electrons,
respectively, $J$ is the total spin obtained by using the appropriate
Clebsch-Gordan coefficients [2,9] and $M$ denotes the magnetic quantum
number of the total angular momentum,

Let us use the principles described in
the previous section and try to find the structure of the
helium atom ground state. Thus, due to the triangular rule (see [9], p. 98)
and in order to be
consistent with $J=0$, we must use configurations where $j_1=j_2$.
Similarly, in order to have an even total parity, we must use configurations
where the two electrons have the same parity.
Thus, the required Hilbert space contains functions of the following form
\begin{equation}
\chi({\bf r}_1,{\bf r}_2) =
F_i(r_1)F_k(r_2)(j^\pi j^\pi 00),
\label{eq:J1J200}
\end{equation}
where $j$ is a positive number of the form $j=n+1/2$, $n$ is an integer
and $\pi = \pm 1$.

The angular parts of any two different functions of $(\!\!~\ref{eq:J1J200})$
are orthogonal. Hence, off-diagonal matrix elements of any pure radial operator
vanish. Since the following discussion is focused
on finding off-diagonal matrix elements of the Hamiltonian, radial
coordinates and radial operators are not always shown explicitly in
expressions.

At this point one can use a given Hamiltonian and construct its matrix.
Before doing this assignment one has to find a practical procedure that
can be used for overcoming the infinite number of configurations that
can be obtained from the different values of $n,\,j$ and $\pi $. For this
purpose one organizes the configurations of $(\!\!~\ref{eq:J1J200})$
in an ascending order of $j$ and
examines a Hilbert subspace made of the first
$N_0$ functions, where $N_0$ is a positive
integer. Here a finite Hamiltonian matrix is obtained and one can
diagonalize it, find the smallest eigenvalue $E_0$
and its associated eigenfunction $\Psi _0$.
The quantities found here represent an approximation for the required
solution. Let this approximate solution be denoted in this form
\begin{equation}
\{E_0,\Psi _0\}.
\label{eq:S0}
\end{equation}
In order to evaluate the goodness of this approximation,
one replaces $N_0$ by $N_1 = N_0 + 1$ and repeats the procedure. The new
solution $\{E_1,\Psi _1\}$
is a better approximation because it relies on a larger Hilbert subspace.
The difference between these solutions
provides an estimate for the goodness of the solutions obtained. This
procedure can be repeated for an increasing value of $N_i$. Thus,
if a satisfactory approximation is reached for
a certain value of $N_i$ then
one may terminate the calculation and use the solution obtained from
this procedure as a good approximation to the accurate solution.

Now we are ready to examine the Hamiltonian's matrix elements.
This examination demonstrates
the advantage of using configurations as a basis
for the Hilbert space. Thus, the angular part of the kinetic energy
of each electron takes the form found for the hydrogen atom and
only diagonal matrix elements do not vanish. The same result is
obtained for the spherically symmetric radial potential
operator $Ze^2/r$ of the nucleus.
It follows that off-diagonal matrix elements can be obtained only
from the interaction between the two electrons. (This quantity does
not exist for the one electron hydrogen atom and for this
reason, each of the hydrogen atom eigenfunctions takes the form of a
unique configuration.) In a full relativistic case the two-electron
interaction takes the form of Breit interaction (see [10], p. 170),
which contains the instantaneous ordinary Coulomb term and a
velocity-dependent term. The existence and the results of the Hamiltonian's
off-diagonal matrix elements are the main objective of this discussion
and it is shown below that for this purpose the examination of the relatively
simple Coulomb term is enough.

Thus, one has to write the $1/r_{12}$ operator in a form that is
suitable for a calculation that uses on the single particle
independent variables
${\bf r}_1,{\bf r}_2$ of the configurations $(\!\!~\ref{eq:J1J200})$.
This objective is achieved
by carrying out a tensor expansion of the interaction (see [9], p. 208).
For the specific case of the Coulomb interaction, the required expression is
(see [11], p. 114)
\begin{equation}
\frac {1}{r_{12}} = \sum _{k=0}^\infty \;\frac {r_<^k}{r_>^{k+1}}
P_k(\cos \theta _{12}).
\label{eq:R12}
\end{equation}
Here $r_<$ and $r_>$ denote the smaller and the larger values of
$r_1$ and $r_2$, respectively and $\theta _{12}$ is the angle between
them. $P_k(\cos \theta _{12})$ is the Legendre polynomial
of order $k$. At this point one uses
the addition theorem for spherical harmonics (see [9], p. 113)
\begin{equation}
P_k(\cos \theta _{12}) = \frac {4\pi}{2k + 1}\sum _{m=-k}^k
(-1)^mY_{k,-m}(\theta_1,\phi _1) Y_{k,m}(\theta_2,\phi _2)
\label{eq:ADDITION_TH}
\end{equation}
and obtains an expansion of the Legendre polynomial $P_k(\cos \theta _{12})$
of $(\!\!~\ref{eq:R12})$
in terms of spherical harmonics that depend on single particle angular
variables. This analysis shows how matrix
elements can be obtained for a Hilbert space whose basis is made
of functions that are an appropriate set of configurations.

At this point the wave functions of the Hilbert space basis
as well as the Hamiltonian operator depend on the radial and the angular
coordinates of single particle functions. The main objective
of this section is to explain why
the electronic states are described as a
linear combination of configurations. It is shown above that the
configurations of the Hilbert space basis are eigenfunctions of the
operators representing the
kinetic energy and the interaction with the spherically symmetric
potential of the nucleus.
Hence, the discussion is limited
to the two particle operator $(\!\!~\ref{eq:R12})$ that depends on
the expansion $(\!\!~\ref{eq:ADDITION_TH})$.

Let us find, for example,
the off-diagonal matrix element of the configurations
$((1{\textstyle \frac {1}{2}}^+)^200)$ and
$((2{\textstyle \frac {3}{2}}^-)^200)$ of the Hilbert space basis
$(\!\!~\ref{eq:J1J200})$. (As explained above, the radial coordinates
do not contribute to the interaction between different configurations.)
The discussion examines the 2-electron Coulomb interaction obtained for the
upper (large) component of the Dirac spinor. Thus,
${\textstyle \frac {1}{2}}^+$ is a spatial s-wave and
${\textstyle \frac {3}{2}}^-$ is a spatial p-wave.
The Wigner-Racah algebra provides explicit formulas for expressions
that depend on the angular coordinates.
Now, as stated above, the main objective of the discussion is to
show that off-diagonal matrix elements do not vanish. For this purpose,
only the main points of the calculation are written and readers can use
explicit reference for working out the details.

The formal form of the angular component of the off-diagonal
matrix element is
\begin{equation}
H_{ij} = <j_1j_2JM|\frac {1}{r_{12}}|j'_1j'_2JM>.
\label{eq:HIJ}
\end{equation}
Here $j_1,\,j_2$ of the ket are
angular momentum values of the first and the second electron,
respectively and they are coupled to a total $J,\,M$. The bra has the
same structure. In the particular case discussed here $J=M=0$ and
$(\!\!~\ref{eq:HIJ})$ takes the form
\begin{equation}
H_{ij} = <{\textstyle \frac {1}{2}}{\textstyle \frac {1}{2}}00
|\frac {1}{r_{12}}|
{\textstyle \frac {3}{2}}{\textstyle \frac {3}{2}}00>.
\label{eq:H01}
\end{equation}

The following points describe the steps used in the calculation
of $(\!\!~\ref{eq:H01})$.

\begin{itemize}
\item[{1.}]
The Wigner-Eckart theorem shows that $(\!\!~\ref{eq:HIJ})$ can be cast
into a product of a {\em Wigner 3j symbol} and a {\em reduced
matrix element} (see [5], p. 117).
\item[{2.}]
In $(\!\!~\ref{eq:R12})$, the expansion
$(\!\!~\ref{eq:ADDITION_TH})$ of $1/r_{12}$ is a {\em scalar product of two
tensors} (see [9], p. 128).
\item[{3.}]
The reduced matrix element of such a scalar product can be put in
the form of a product of a {\em Racah coefficient} and two reduced
matrix elements that depend on the first and the second electron,
respectively (see [5], p. 129).
\item[{4.}]
Each of these reduces matrix elements takes the form $<slj||Y_k||sl'j'>$
where $sl$ denote single particle spin and spatial angular momentum
that are coupled to the particle's total angular momentum $j$.
In the specific case discussed here it is
$<{\textstyle \frac {1}{2}}0{\textstyle \frac {1}{2}}||Y_1||
{\textstyle \frac {1}{2}}1{\textstyle \frac {3}{2}}>$.
The value of the last expression can be readily obtained as
a product of a square root of an integer and a Wigner $3j$ symbol
(see [9], p. 521). The final value is
\begin{equation}
<{\textstyle \frac {1}{2}}0{\textstyle \frac {1}{2}}||Y_1||
{\textstyle \frac {1}{2}}1{\textstyle \frac {3}{2}}> =
\frac {-2}{\sqrt {4\pi}}.
\label{eq:RED_MAT_ELEM}
\end{equation}

\end{itemize}

This discussion shows that
the Hamiltonian's off diagonal matrix elements do not vanish
for the $J=0$ ground state of the He atom. It means
that a single configuration does not describe accurately this state.
The next step is to carry out an explicit calculation and find out
how good is the usage of a single configuration. This task has
already been carried out [4] and it was proved that the description
of the ground state of the He atom requires many configurations. Here
radial and angular excitations take place and no single
configuration plays a dominant role.

\vglue 0.66666in
\noindent
{\bf 4. Discussion}
\vglue 0.33333in

Several aspects of the conclusion obtained in the previous section
are discussed below.

Intuitively, the multiconfiguration structure of the ground state
may be regarded as a mistake. Indeed, the ground state takes the
lowest energy possible. Hence, how can a mixture of a lower
energy state and a higher energy state yield a combined state whose energy
is lower than either of the two single mono-configuration states?
The answer to this question relies on a solid mathematical basis. Thus,
a diagonalization of a Hermitian matrix reduces the lowest
eigenvalue and increases the highest eigenvalue (see e.g. [12], pp. 420-423).
Hence, {\em for a Hermitian matrix,
any off-diagonal matrix element increases the difference
between the corresponding diagonal elements}. It means that the
smaller diagonal element decreases and the larger diagonal element
increases. Since the Hamiltonian is a Hermitian operator, one concludes
that if the Hilbert space basis yields a
non-diagonal Hamiltonian matrix then the lowest eigenvalue "favors"
eigenfunctions that are a linear combination of the Hilbert space
basis functions.

It is shown in the previous section that the non-vanishing off-diagonal
matrix elements stem from the two body Coulomb interaction between
electrons. Thus, the tensor expansion of the interaction
$(\!\!~\ref{eq:R12})$ casts the 2-body Coulomb interaction into
a series of Legendre polynomials where $cos \theta _{12}$ is
the polynomial's argument. Evidently, any physically meaningful
interaction depends on the distance between the interacting particles. Hence,
an expansion in terms of the Legendre polynomials can be obtained.
This expansion proves that
the mathematical procedure described in the previous section
has a comprehensive validity (see [9], p. 208). Thus, what is found
in the previous section for electrons in the He atom ground state
also holds for quarks in the proton. Moreover, the proton
is an extremely relativistic system of quarks and, as such, its
spin-dependent interactions are expected to be quite strong.
Evidently, spin dependent interactions make a contribution to
off-diagonal matrix elements. On the basis of this
conclusion, one infers that the proton's quark state must be described by a
linear combination of many configurations.

A polarized
proton experiment has been carried out where the instantaneous spin
direction of quarks was measured [6]. The measurements have shown that
{\em the total quark spin constitutes a rather small fraction
of the proton's spin.} This result is in a complete agreement with
the mathematical analysis carried out above. Thus, the relativistic
proton dynamics indicates that the $jj$-coupling provides a better
approach (and this is the reason for the usage of this notation here).
In each quark configuration, spin and spatial angular
momentum are coupled to a total single particle $j$-value and the
Clebsch-Gordan coefficients determine the portion of spin-up and
spin-down of the quark. Next, The relativistic quark state indicates
that, unlike the case of the hydrogen atom, the lower part of the
Dirac spinor of quarks is quite large. As is well known, if in the
upper part of a Dirac spinor is $l=j\pm 1/2$ then its lower part
is $l=j\mp 1/2$. Hence, different Clebsch-Gordan coefficients are
used for the upper and the lower parts of the Dirac spinor. Furthermore,
in different configurations, different Clebsch-Gordan coefficients
are used for the single particle coupling of the three quarks
to the total proton's spin
and the overall weight of the spin-up and spin-down components
takes a similar value. This argument indicates that the outcome
of [6] is quite obvious and that if the experiment would have
yielded a different
conclusion where {\em quarks carry the entire proton's spin then
this result should have been regarded as a real crisis of fundamental
quantum mechanical principles.} This discussion also shows that the quite
frequently used description of the results of [6] as ``the proton spin crisis"
is unjustified.

Computers are based on quantum mechanical processes that take
place in solid state devices. Hence, it is clear that people who
have established the laws of quantum mechanics had no access to
the computational power
of computers. For this reason, several approximations have been
contrived in order to get an insight into atomic structure.
A method that deals with configurations is called {\em central
field approximation} (see [13], p. 225). Here, for every electron,
the actual field of all other electrons is replaced by an approximate
spherically symmetric
radial field. Evidently, as explained in the third section, such a
radial field does not cause a configuration mixture and, in this
approximation, a single configuration is used for describing atomic
states. This approach is frequently used in a description of the
Mendeleev's periodic table (see [13], pp. 240-247).

However, even in the early days of quantum mechanics,
the central field approximation has been regarded as an
approximation and people have constructed mathematical tools for
treating the multi-configuration atomic structure which is known
as the Wigner-Racah algebra of angular momentum. These mathematical
tools have been used in the early days of electronic computers [4]
and the result is quite clear: {\em many configurations are required
even for the simplest case of the
ground state $J=0$ of the 2-electron He atom and no single configuration
plays a dominant role.} Today, this outcome is still known (see [11], p. 116)
but unfortunately not widely known. Thus, [11] is based on lectures
delivered in a chemistry department. On the other hand,
the birth and the long duration of the idea concerning
{\em the proton spin crisis} prove that
this fundamental physical issue is indeed not widely known. This paper
has been written for the purpose of improving the present status.

\newpage
References:
\begin{itemize}
\item[{*}] Email: elicomay@post.tau.ac.il  \\
\hspace{0.5cm}
           Internet site: http://www.tau.ac.il/$\sim $elicomay

\item[{[1]}] E. A. Poe, {\em The Raven} (1845).
\item[{[2]}] E. U. Condon and G. H. Shortley, {\em The Theory of
Atomic Spectra}, (University Press, Cambridge, 1964).
\item[{[3]}] G. R. Taylor and R.G. Parr, Proc. Natl. Acad. Sci. USA
{\bf 38}, 154 (1952).
\item[{[4]}] A. W. Weiss, Phys. Rev. {\bf 122}, 1826 (1961).
\item[{[5]}] Textbooks use the central field approximation and do not
analyze the goodness of this argument. [11] (see p. 116)
is a notable exception.
\item[{[6]}] J Ashman et al. (EMC) Phys. Lett. {\bf B206}, 364 (1988).
\item[{[7]}] E. P. Wigner, Annals of Math., {\bf 40}, 149 (1939).
\item[{[8]}] S. S. Schweber, {\em An Introduction to Relativistic
Quantum Field Theory}, (Harper \& Row, New York, 1964). Pp. 44-53.
\item[{[9]}] A. de-Shalit and I. Talmi, {\em Nuclear Shell Theory}
(Academic, New York, 1963).
\item[{[10]}] H. A. Bethe and E. E. Salpeter, {\em Quantum Mechanics
of One and Two-Electron Atoms} (Springer, Berlin 1957).
\item[{[11]}] R. N. Zare, {\em Angular Momentum} (Wiley, New York, 1988).
\item[{[12]}] C. Cohen-Tannoudji, B. Diu and F. Laloe,
{\em Quantum Mechanics} Vol. 1 (Wiley, New York,2005).
\item[{[13]}] L. D. Landau and E. M. Lifshitz {\em Quantum Mechanics}
(Pergamon, London, 1959).

\end{itemize}

\end{document}